# Sessile drops in microgravity

**Amelia Carolina Sparavigna**
**Dipartimento di Scienza Applicata e Tecnologia**
**Politecnico di Torino, Torino, Italy**

*Interfaces with a liquid are governing several phenomena. For instance, these interfaces are giving the shape of sessile droplets and rule the spread of liquids on surfaces. Here we analyze the shape of sessile axisymmetric drops and how it is depending on the gravity, obtaining results in agreement with experimental observations under conditions of microgravity.*

Many phenomena can be observed when we have a system containing some interfaces with a liquid. We can see for instance that sessile drops tend to have a lenticular shape and water spreads on some surfaces whereas it is remaining in isolated drops on others. These phenomena can be understood through the concept of surface tension. From a molecular point of view, we can imagine a molecule and the forces acting on it. If the molecule is at the surface of a liquid, the net force on it due is different, compared to that acting on a particle in the bulk. In fact, the net force acting on a molecule in the bulk is null in average, and so no net force exists pulling the molecule in any given direction. A molecule at the surface, however, will feel an unbalanced force due to the lack of near neighbors in the direction of the vapor outside the liquid. The result is that a molecule at the surface be pulled into the bulk.

In order to study the surface tension and interfacial forces between liquids and solid substrates, the analysis of the shape of sessile drops is one of the most used methods. There are several reasons supporting the preference for this method, in particular it is easy to apply and requires just a small quantity of liquid. [1] Today, the method is combined with the use of image processing and a numerical computation on personal computers to fit data. When applied to axisymmetric droplets, the overall procedure is defined as ADSA, Axisymmetric Drop Shape Analysis [2,3]. The calculation of the shape of the droplets is based on the numerical solution of the Young-Laplace equation. Here we discuss the sessile droplets according to ADSA, and how their shape depends on the gravity to compare it with the experimental observations obtained in conditions of microgravity.

**Microgravity and interfacial phenomena**
As a matter of fact, surfaces and interfaces need more investigation. For instance, a better understanding of interfacial phenomena such as wetting will help improving materials processing [4]. Moreover, interfaces dominate the properties and behavior of advanced composite materials [5]. However, the control of wetting and spreading poses both scientific and technological challenges. This is why many efforts have been devoted to study how the wetting determines the features of fluid interfaces in a regime of microgravity too. This peculiar environment provides scientists with an excellent opportunity to study all those features of wetting and surface tensions that are normally masked by the Earth's gravity [4].

The term microgravity, indicated by the symbol μg, is more or less a synonym of weightlessness, and indicates a gravitational force which is very small. Then, the best laboratory to investigate microgravity is the International Space Station [6]. Another possibility is to use airplanes in a parabolic flight or the microgravity drop towers. These are structures used to produce a controlled period of weightlessness. Air bags and magnetic or mechanical brakes are used to arrest the fall of the experimental payload.

The period of weightless is of about a few minutes, that is, when the sample is freely falling in the tube. During its flight the sample can be characterized with instruments. In fact, the first drop tower was the Leaning Tower of Pisa that it is popularly supposed Galileo used to demonstrate that bodies having different mass fall with the same acceleration [7].

The research of interfacial phenomena in microgravity focuses on how an interface acquires and maintains its shape [8-10]. For instance, we can see that reducing the gravity droplets assume a spherical shape (see Fig.1).

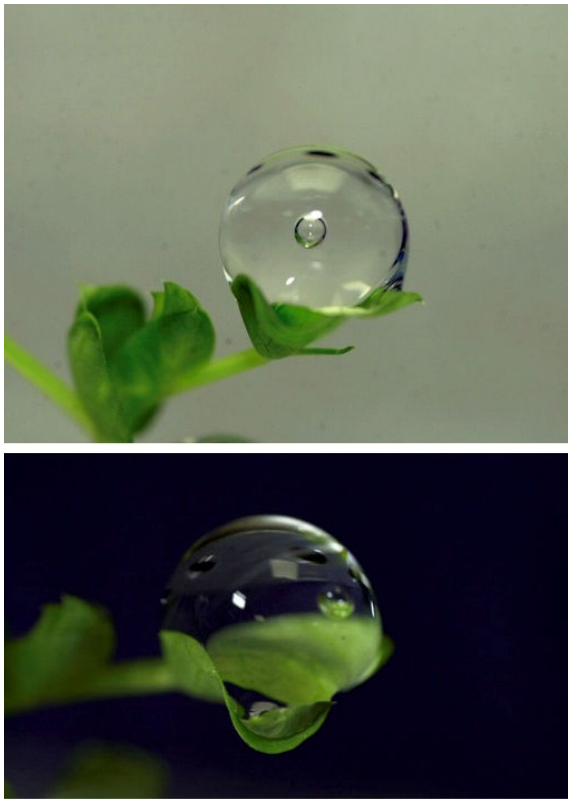

**Figure 1: Droplets and bubbles in the space garden of the International Space Station (Courtesy: NASA).**

Besides the analysis of static surfaces, the dynamics is studied in response to heating, cooling, and chemical influences. On the ground, the gravity causes buoyancy, which is generating convection. In the space, buoyancy disappears; however a different type of convection exists, originated by the surface tension, convection which is depending on temperature and, in the case of a mixture, on concentration. Accordingly, if a temperature or concentration distribution exists, a distribution of the surface tension also arises, with flows occurring towards the places where the surface tension is higher. This mechanism is the Marangoni convection [11]. It is therefore important to study the surface tension in microgravity conditions, because it has a great impact in the micro-world where the gravity becomes less significant with decreasing scale, and therefore, this study can help improving the micro-fluid technology on the ground [11].

**Contact angle and wetting**

When a drop of liquid is placed on a solid surface, a triple linear interface is formed where solid, liquid and vapor phases coexist. This triple line will move in response to the forces arising from the three interfacial tensions until an equilibrium position is established. In Figure 2 we see a drop of liquid ($L$) on a solid surface ($S$) with vapor ($V$) as the third phase. The angle θ between the solid surface and the tangent to the liquid surface at the line of contact with the solid is known as the contact angle. At equilibrium, tensions will be in balance and thus: $\gamma_{VS} = \gamma_{LS} + \gamma_{VL} \cos\theta$. The origin of the contact angle is therefore in the cohesive forces between molecules in the liquid drop and the adhesive forces between liquid molecules and the surface.

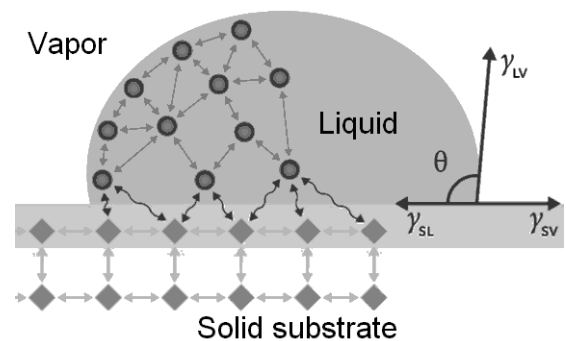

**Figure 2: A molecule at the surface of a drop is subjected to a net force which is the origin of the surface tension. On the right of the drop, the contact angle is shown.**

We can see then that some surfaces have a very high contact angle for water, while for others it is so low that we cannot measure it [12]. A surface having a good affinity for water has a low contact angle. Such a surface is called hydrophilic. If a surface has preponderantly non-polar groups, such as a polymer surface, the surface is hydrophobic and the contact angle will be large. Wetting is determined by the equilibrium contact angle too. If θ < 90°, the liquid is said to wet the solid; if θ = 0, there is complete or perfect wetting; if θ > 90°, the liquid does not wet the solid [13]. However, such features of a surface can be controlled using some specific treatments, such as atmospheric plasma [14,15].

The measurement of contact angle is therefore a simple method to gain qualitative information about the surface. Many surfaces display an apparent hysteresis, giving different values of the contact angle when the measurement is obtained from a drop of increasing size (advancing contact angle) or of diminishing size (receding contact angle). A cause of this hysteresis is in the roughness of the surface, and then the difference between advancing and receding contact angles is used to obtain information about the nature of the surface [16].

**The Young-Laplace equation and its solution**

ADSA methods are based on the numerical fit between the shape of the drops as experimentally evidenced and the solution of the mathematical model given by the Laplace-Young equation for capillarity [2].

The general form of the Laplace-Young equation describes the capillary pressure difference sustained across an interface, due to the surface tension. In particular, the equation relates the pressure difference to the shape of the surface:

$$\Delta p = \gamma \left( \frac{1}{R_1} + \frac{1}{R_2} \right) \quad (1)$$

where $R_1, R_2$ are the principal radii of curvature and $\gamma$ the surface tension. The change of pressure is then given by two contributions, the change in hydrostatic pressure and the change in pressure due to the curvature of the drop. However, the sessile drops have not a constant curvature, and then Eq.1 needs a local coordinate system, which is the spherical coordinate one. Figure 3 illustrates the coordinate system with origin located at the apex of the drop [1-3].

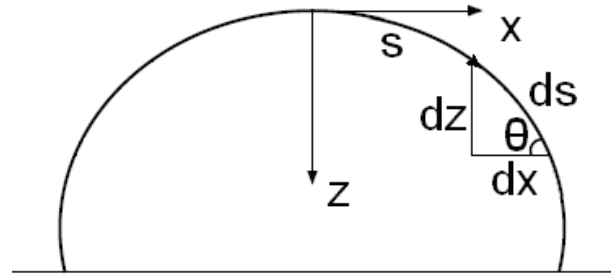

**Figure 3:** Spherical coordinate system used aiming to describe the local curvature of the sessile drop.

The model and the numerical method to determine the shape of the sessile drops are given in Ref.2. Equations are:

$$\frac{dx}{ds} = \cos\theta$$
$$\frac{dz}{ds} = \sin\theta \quad (2)$$
$$\frac{d\theta}{ds} = 2b + cz - \frac{\sin\theta}{x}$$

In Eq.2, $s$ is the arc length, $b$ and $c$ are the curvature at the origin of the coordinates and the capillarity constant of the system ($g \cdot \Delta\rho / \gamma$), respectively.

To apply the model (2) investigating the shape of droplets in microgravity, I used a Runge-Kutta method of solution in a Fortran program, with four stages. For instance, the second equation in formulae (2) can be solved using the third one, in the following manner. Let us choose a fix small value of $p=ds$, the increment of arc length, and values $\theta_0$, $z_0$, $r_0$, to initialize the iteration. Then:

$ps_0 = (2 \cdot b + c \cdot z_0 - r_0)$
$\theta_1 = \theta_0 + ps_0 \cdot p$
$dz_1 = \sin(\theta_1) \cdot ps_0$
$ps_1 = (2 \cdot b + c \cdot (z_0 + dz_1) - r_0)$

```
do i=2,4
θ_i = θ_0 + ps_{i-1}·p/2.
dz_i = sin(θi)·p/2.
ps_i = (2.·b + c·(z_0+dz_i) − r_0)
end do
```

so having the increment $dz$. The same approach is used to solve the first equation (2), and to evaluate the volume of the drops.

The capillarity constant in the regime on normal gravity, is assumed as 13.5 cm$^{-2}$, according to Ref.2. In order to compare the effect of microgravity, for which I used a value of $10^{-6}$g, let us consider drops having a fixed volume (1 cm$^3$), for several apex curvatures.

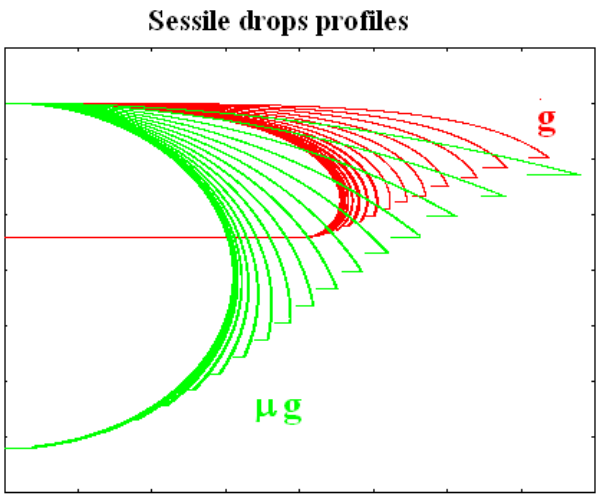

**Figure 4: Profile of sessile drops, having the same volume but different apex curvatures, in the case of gravity (red) and microgravity (green). Note the**

The profile of droplets in the coordinate system of Figure 3 is given in the Figure 4. We have, in red, the profiles of droplets in the normal gravity field, and, in green, those in microgravity. The droplets have the same fixed volume. We can see several droplets having different contact angles with the horizontal surface, represented by a short horizontal line (the droplets are displayed as having the same apex).

We can see that the shape in normal gravity is more flat than that of a droplet, having besides the same volume the same contact angle too, in condition of microgravity. The shape can be spherical. And this is what is observed in experiments [8].

**Conclusions**

Reference 8 was published in 2012, and is presenting an initial work aimed to the development of a database of contact angles of sessile drops in reduced gravity. According to the authors, there is not such a database, the creation of which is fundamental for investigating the role of heat and mass transfer in reduced gravity and future engineering designs. In fact, this is in agreement with Ref.11, which is remarking the importance of studies in microgravity, the results of which can be applied to micro- and mesoscopic systems where the role of gravity is reduced.

In Ref.8, the authors investigated the water and ethanol on PTFE and aluminum surfaces. They observed that the droplet shape is adequately described by the Young-Laplace equation in reduced gravity environments. A variation is occurring in the diameter and height upon the surface, to obtain a sphere-like volume. The dimensional variations showed that an increase in the Bond number $B$, $(B=(r/L_c)^2 = g \cdot \Delta\rho \cdot r^2 / \gamma$, where $r$ is the wetting radius and $L_c$ is the capillary length) correlated well with the change in the wetted diameter and the height of the droplet. In fact, the triple line perimeter of the drops is lower in reduced gravity than in normal gravity. According to [8] then, the Young-Laplace equation can be used to determine the contact angle in reduced gravity for small droplets, but it is not optimal to describe the contact angle for larger drops. Moreover, hysteresis exists because of drop pinning due to the roughness of the surface, and the pinning can be affected by gravity too. Therefore Ref.8 is claiming that more experiments are necessary to understand the behavior of large droplets to improve the theory.

Here I have proposed a short discussion of the problem suitable for teaching purposes; in particular to show an application of the Young-Laplace equation, application which is involving image processing and numerical methods, besides being important for industrial problems. The aim is that of stimulate the students towards the study of the problems connected with a microgravity environment.

**References**

1. Mina Hoorfar and A. Wilhel Neumann, Axisymmetric Drop Shape Analysis (ADSA), Applied Surface Thermodynamics, Second Edition, Yi Zuo Editor, CRC Press, 2010, Pages 107-174.
2. O.I. del Río and A. W. Neumann, Axisymmetric Drop Shape Analysis: Computational Methods for the Measurement of Interfacial Properties from the Shape and Dimensions of Pendant and Sessile Drops, Journal of Colloid and Interface Science, 196, 136–147 (1997).
3. Mina Hoorfar, Development of a PC version for axisymmetric drop shape analysis (ADSA), 2001, thesis, Department of Mechanical and Industrial Engineering University of Toronto
4. Microgravity Science Primer, A Teacher's Guide with Activities in Science, Mathematics, and Technology, EG-1997-08-110-HQ, NASA, available at the web page http://www.nasa.gov/pdf/315962main_Microgravity_Science_Primer.pdf
5. Marcello Lappa, Fluids, Materials And Microgravity: Numerical Techniques And Insights Into Physics, Elsevier, 2004
6. Unlocking Mysteries in Microgravity, NASA FActs, National Aeronautics and Space Administration, Glenn Research CenterFS-1999-07-007-GRC
7. Carl G. Adler, Byron L. Coulter, Galileo and the Tower of Pisa experiment, merican Journal of Physics, March 1978, Volume 46, Issue 3, pp. 199
8. Antoine Diana, Martin Castillo, David Brutin, Ted Steinberg, Sessile Drop Wettability in Normal and Reduced Gravity, Microgravity Sci. Technol., 24, 195-202, Springer, 2012, DOI 10.1007/s12217-011-9295-0
9. Zhi-Qiang Zhu, David Brutin, Qiu-Sheng Liu, Yang Wang, Alexandre Mourembles, Jing-Chang Xie, Lounes Tadrist, Experimental Investigation of Pendant and Sessile Drops in Microgravity, Microgravity Sci. Technol. (2010) 22:339–345, DOI 10.1007/s12217-010-9224-7
10. David Brutin, ZhiQuiang Zhu, Ouamar Rahli, JingChang Xie, QuiSheng Liu, Lounes Tadrist, Sessile Drop in Microgravity: Creation, Contact Angle and Interface, Microgravity Sci. Technol, 21, S67-s76, Springer, 2009, DOI 10.1007/s12217-009-9132-x
11. Hiroshi Kawamura, Fluid Science under Microgravity, JAXA, Japan Aerospace Exploration Agency, http://www.jaxa.jp/article/special/kibo/kawamura_e.html
12. S.W. Rienstra, The shape of a sessile drop for small and large surface tension, Journal of Engineering Mathematics, 24, 193-202, 1990.
13. D. L. Schodek, P. Ferreira, and M. F. Ashby, Nanomaterials, Nano-technologies and Design: An Introduction for Engineers and Architect, Butterworth-Heinemann, 24/mar/2009, Pag.409.
14. R. Wolf, A.C. Sparavigna, Role of Plasma Surface Treatments on Wetting and Adhesion, Engineering, vol. 2(6), pp.397-402, 2010, DOI: 10.4236/eng.2010.26052
15. A.C. Sparavigna, Plasma treatment advantages for textiles, arXiv, 2008, arXiv:0801.3727 [physics.pop-ph]
16. Russell Stacy, Contact Angle Measurement Technique for Rough Surfaces, Thesis, Michigan Technological University, 2009